\documentclass[%
 reprint,
 amsmath,amssymb,
 aps,
]{revtex4-1}

\usepackage{graphicx}
\usepackage{dcolumn}
\usepackage{bm}
\usepackage{xspace}
\usepackage{lineno,hyperref}
\usepackage{xcolor}

\newcommand{\pt}{\ensuremath{p_{\rm T}}\xspace}

\newcommand{\nch}{\ensuremath{N_{\rm ch}}\xspace}
\newcommand{\nmpi}{\ensuremath{N_{\rm mpi}}\xspace}

\begin{document}

\preprint{APS/123-QED}

\title{Extraction of the multiplicity dependence of Multiparton Interactions from LHC pp data using Machine Learning techniques}
\author{Antonio Ortiz}
 \email{antonio.ortiz@nucleares.unam.mx}
\author{Erik Zepeda}%
 \email{eazg@ciencias.unam.mx}
\affiliation{%
Instituto de Ciencias Nucleares, Universidad Nacional Aut\'onoma de M\'exico,\\
 Apartado Postal 70-543, M\'exico Distrito Federal 04510, M\'exico 
}%

\date{\today}

\begin{abstract}

Over the last years, Machine Learning (ML) methods have been successfully applied to a wealth of problems in high-energy physics. For instance, in a previous work we have reported that using ML techniques one can extract the Multiparton Interactions (MPI) activity from minimum-bias pp data. Using the available LHC data on transverse momentum spectra as a function of multiplicity, we reported the average number of MPI ($\langle N_{\rm mpi} \rangle$) for minimum-bias pp collisions at $\sqrt{s}=5.02$ and 13\,TeV. In this work, we apply the same analysis to a new set of data. We report that $\langle N_{\rm mpi} \rangle$ amounts to $3.98 \pm 1.01$ for minimum-bias pp collisions at $\sqrt{s}=7$\,TeV. These complementary results suggest a modest center-of-mass energy dependence of $\langle N_{\rm mpi} \rangle$. The study is further extended aimed at extracting the multiplicity dependence of $\langle N_{\rm mpi} \rangle$ for the three center-of-mass energies. We show that our results qualitatively agree with existing ALICE measurements sensitive to MPI. Namely, $\langle N_{\rm mpi} \rangle$ increases approximately linearly with the charged-particle multiplicity. But, it deviates from the linear dependence at large charged-particle multiplicities. The deviation from the linear trend can be explained in terms of a bias towards harder processes given the multiplicity selection at mid-pseudorapidity. The results reported in this paper provide additional evidence of the presence of MPI in pp collisions, and they can be useful for a better understanding of the heavy-ion-like behaviour observed in pp data.

\end{abstract}

\maketitle


\section{Introduction}

The possibility of having Multiparton Interactions (MPI), i.e. several parton-parton interactions within the same hadron-hadron collision, is expected given the composite nature of hadrons. Indeed, at the Large Hadron Collider (LHC) energies, already at a transverse momentum transfer of a few GeV/$c$ the cross section for leading order (LO) parton-parton scatterings exceeds the total pp inelastic cross section~\cite{Bahr:2008wk}.  This apparent inconsistency can be resolved by considering additional partonic scatterings within the same pp collision~\cite{Sjostrand:1986ep}. Data support the presence of MPI in pp collisions ( e.g.~\cite{Abelev:2012sk},~\cite{Abelev:2013sqa},~\cite{Ortiz:2017jho},~\cite{Ortiz:2020rwg} ). For instance, in pp collisions at $\sqrt{s}<0.2$\,TeV the evolution of the charged-particle multiplicity distribution as a function of $\sqrt{s}$ follows the Koba-Nielsen-Oleson (KNO) scaling with scaling variable $z=\nch / \langle \nch \rangle$~\cite{Koba:1972ng}.  However, such a scaling is violated at higher energies~\cite{Alner:1985wj}. This effect can been interpreted as a consequence of particle production through MPI~\cite{Dremin:2011sa}.   

Beyond the importance of Multiparton Interactions for high-energy physics, the study of its effects in pp collisions has recently attracted the attention of the heavy-ion community (see e.g.~\cite{Blok:2017pui,Blok:2018xes}). This is because the multiplicity dependent studies of pp data unveiled heavy-ion-like features, i.e.  azimuthal anisotropies~\cite{Khachatryan:2010gv}, the enhancement of (multi-)strange hadrons~\cite{ALICE:2017jyt}, as well as the mass ordering in the hadron \pt spectra~\cite{Acharya:2018orn}. Besides the hydrodynamical approach~\cite{Bozek:2011if,Nagle:2018nvi}, Multiparton Interactions, which is a key mechanism of Monte Carlo (MC) generators like Pythia~8~\cite{Sjostrand:2014zea} and Herwig~7~\cite{Gieseke:2012ft}, offer an alternative possibility to explain the observed phenomena. For instance, color reconnection and MPI can mimic radial flow patterns in pp collisions~\cite{Ortiz:2013yxa}. Models based on the QCD theory of MPI have been shown to explain collectivity from interference effects in hadronic collisions with \nmpi parton-parton scatterings~\cite{Blok:2017pui,Blok:2018xes}. Pythia~8 with rope hadronization model~\cite{Bierlich:2015rha}, which assumes the formation of ropes due to overlapping of strings in a high-multiplicity environment (high \nmpi), describes the strangeness enhancement~\cite{Nayak:2018xip}. Regarding the phenomena at large transverse momentum (\pt), the model also produces some features which are present in heavy-ion data~\cite{Mishra:2018pio,Jacobs:2020ptj,Ortiz:2020dph}. 

It is worth mentioning that the early LHC data already suggested that in high-multiplicity pp collisions, the MPI activity could be more relevant than assumed and that this could give rise to new effects~\cite{Abelev:2012sk}. For this reason we have proposed the extraction of MPI from minimum-bias pp data using Machine Learning (ML) methods~\cite{Ortiz:2020rwg}. In this paper, we extend that study aiming at extracting the multiplicity dependence of \nmpi from the available ALICE data~\cite{Acharya:2019mzb,Acharya:2018orn} at the LHC. Our results are contrasted with ALICE data~\cite{Abelev:2013sqa} at lower energies, and discussed in terms of what we know from the Pythia~8 model. 

The paper is organised as follows:  section 2 describes the analysis, where the input variables and the models used for the study are discussed.  Results are presented  in  section  3,  and  finally  section  4  contains  a summary and outlook.

\section{Analysis}

The goal of our work is the extraction of the average number of MPI from LHC data. Our approach relies on a multivariate regression technique based on Boosted Decision Trees (BDT), which follows the strategy reported in Ref.~\cite{Ortiz:2020rwg}. A regression tree is a binary tree structured regressor in which repeated yes or not decisions are taken on one single variable. In this way, the phase space is split into many regions where each output node represents a specific value of the target variable. For regression tasks the boosting algorithm used is the gradient boost, which tries to minimize the loss-function which describes how the model is predictive with respect to the training data.   The study is conducted using the Toolkit for Multivariate Analysis (TMVA) framework which provides a ROOT-integrated machine learning environment for the processing and parallel evaluation of multivariate classification and regression technique~\cite{Voss:2007jxm}. In our study we consider the following TMVA options: \texttt{NTrees=2000}, \texttt{Shrink-age=0.1}, \texttt{BaggedSampleFraction=0.5}, \texttt{nCuts=20}, \texttt{MaxDepth=4} (details about these options can be found in Ref.~\cite{Voss:2007jxm}). The training is performed using pp collisions at $\sqrt{s}=13$\,TeV simulated with Pythia~8.244~\cite{Sjostrand:2014zea} event generator (tune 4C~\cite{Corke:2010yf}). For the multiplicity dependent studies, we use the same input variables as reported in Ref.~\cite{Ortiz:2020rwg}. The choice of the variables is based on their correlation with \nmpi~\cite{Cuautle:2015fbx}, as well as their availability as published data. We consider the event-by-event average transverse momentum and the mid-pseudorapidity charged particle multiplicity ($N_{\rm ch}$). The charged particle multiplicity is strongly correlated with the MPI activity (see e.g. Ref.~\cite{Cuautle:2015fbx}). Moreover, the correlation between these quantities encodes information about the underlying particle production mechanism. Given the kinematic restrictions from the ALICE data, these quantities are calculated for primary charged particles within $|\eta|<0.8$, in addition, the average \pt considers tracks with transverse momentum above 0.15\,GeV/$c$.

\begin{figure*}
\includegraphics[width=0.85\textwidth]{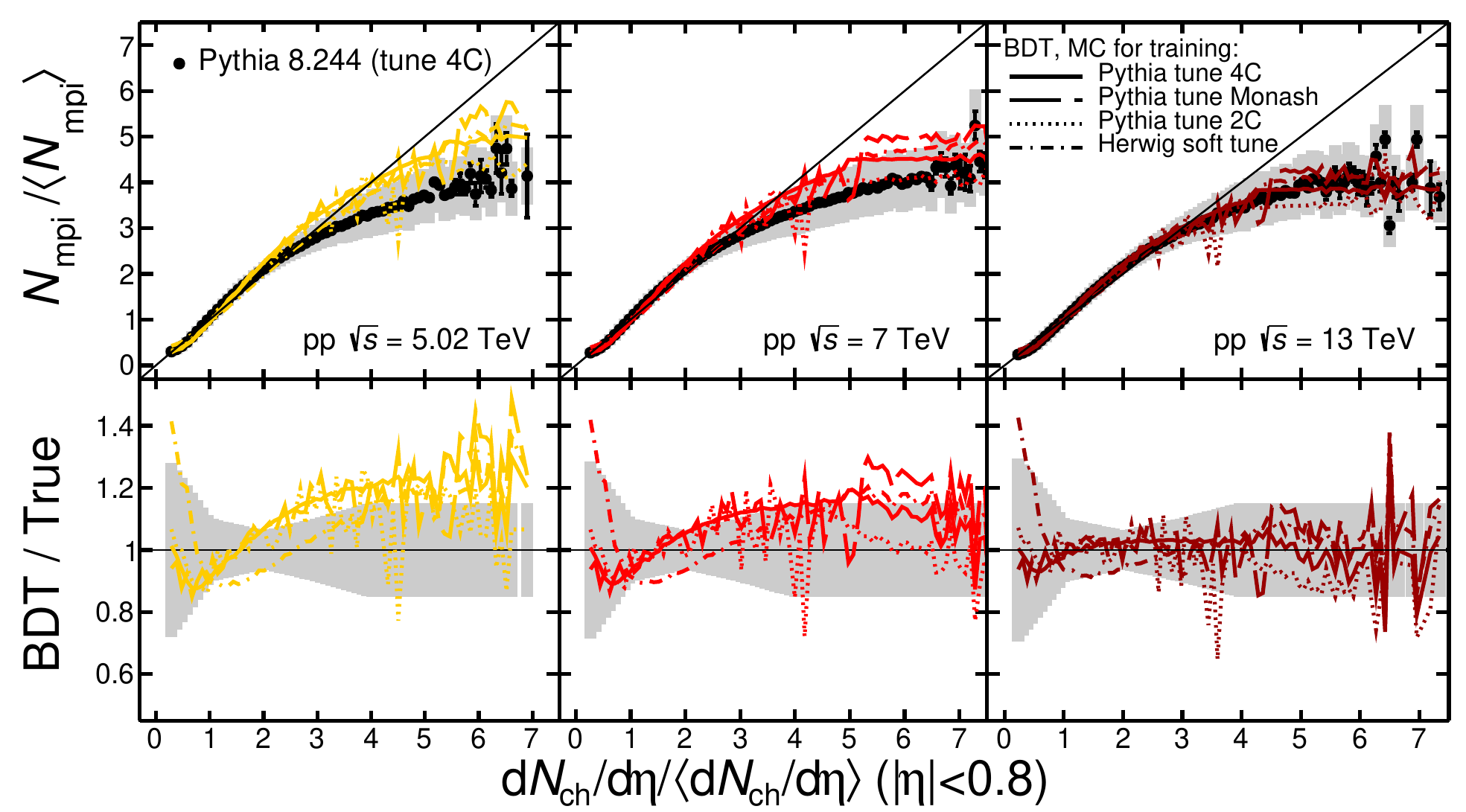}
\caption{Monte Carlo closure test using pp collisions at $\sqrt{s}=5.02$ (left), 7 (middle) and 13\,TeV (right) simulated with Pythia~8 tune 4C. The top panels display the self normalized average number of Multiparton Interactions as a function of the self-normalized mid-pseudorapidity charged particle multiplicity. The Pythia 8 results (solid markers) are shown along BDT results (lines). The results from ML are obtained considering different MC models for training: 4C (solid line), Monash (dashed line), 2C (dotted line) and the soft tune of Herwig 7 (dash-dotted line). Ratios between ML results and the true values given by Pythia are shown in the bottom panels. In all panels the grey band indicates the systematic uncertainties (see the text for more details).}
\label{fig:1}
\end{figure*}

The systematic uncertainty which was assigned in our previous study took into account a variation of the model. To this end, different Pythia~8 tunes were used for training: 2C, 4C and Monash~2013. In the present study, we proceed in the same way. The main features of the Pythia~8 tunes used in our analysis are listed below.

\begin{itemize}

\item The tune 2C was obtained from fits to TEVATRON data, therefore, it was not presented as  a ``complete'' MC tune for LHC~\cite{Corke:2010yf}. Instead, it was provided as a starting point for more adequate tunes using the LHC data. This explains why this tune gives the worst description of the LHC data. We have chosen this model in order to evaluate the impact in our results if BDT are trained with a model which is known to fail to describe the data.

\item The model 4C, on the other hand, used the early LHC minimum-bias and underlying-event data (pp at $\sqrt{s}=0.9$ and 7\,TeV) for tuning, which the model 2C significantly underestimated. 

\item The Monash~2013 model is tuned to a bigger set of LHC data~\cite{Skands:2014pea}. Contrary to the previous tunes, Monash~2013 starts from a more careful tune to LEP data, and it involves several parameter changes. 

\end{itemize}

The effects of the hadronization model used for training is also investigated using the Monte Carlo generator  Herwig~7.2~\cite{Bellm:2019zci} for training instead of Pythia~8. The effects of both the MPI model and hadronization model are considered in the systematic uncertainties.

Before processing the data using the trained BDT, first we show that the procedure is robust against the MC model used for training. To this end, we perform a Monte Carlo closure test. Figure~\ref{fig:1} shows the correlation between the self-normalized number of Multiparton Interactions ($\nmpi / \langle \nmpi \rangle$) and the self-normalized mid-pseudorapidity charged particle multiplicity ($\nch / \langle \nch \rangle$) in pp collisions at $\sqrt{s}=5.02$, 7 and 13\,TeV. The results were obtained using Pythia~8 tune 4C. For $\nch / \langle \nch \rangle<3$, the self normalized \nmpi increases linearly with the event multiplicity. While, for higher multiplicities, we observe a deviation of the self normalized \nmpi with respect to the linear trend. This observation suggests that very high-multiplicity pp collisions can only be produced by high-multiplicity jets~\cite{Ortiz:2016kpz}. The figure also displays the results obtained from regression (lines). Namely, the MC information (average \pt and multiplicity) of pp collisions at $\sqrt{s}=5.02$, 7 and 13\,TeV simulated with Pythia~8 tune 4C, was evaluated using four different sets of BDT. Each one was trained considering different MC models: the three Pythia~8 tunes described above, as well as the soft tune of Herwig~7.2. Figure~\ref{fig:1} shows that using ML-based regression, one can recover the energy and multiplicity dependence. The small variations with respect to the true correlation (markers) are well covered by the systematic uncertainties, which amount to 30\% for $\nch / \langle \nch \rangle \rightarrow 0$, and 15\% for  $\nch / \langle \nch \rangle >4$. It is worth mentioning that other ML methods were also tested, the results were compatible with those obtained using BDT. Therefore, the systematic uncertainty on the BDT approach itself was assumed to be negligible.

Given that event-by-event correlations between  $\langle p_{\rm T} \rangle$ and \nch are not available as public data, a strategy to build them was developed. To this end, we built a toy MC using the available ALICE data~\cite{Acharya:2018orn,Acharya:2019mzb}, which contain the \pt spectra for different multiplicity classes defined by the event activity at either mid-pseudorapidity (tracklets-based estimator) or forward pseudorapidity (V0M-based estimator). The tracklet-based estimator covers the pseudorapidity interval $|\eta|< 0.8$. One tracklet is a track segment defined by pairs of clusters, one cluster in each layer of the Silicon Pixel Detector (SPD) of ALICE~\cite{doi:10.1142/S0217751X14300440}. The V0M estimator is based on the total charge deposited in the forward detector covering the pseudorapidity regions $2.8 < \eta < 5.1$ and $-3.7 < \eta < -1.7$.

\begin{figure}[t]
\includegraphics[width=0.45\textwidth]{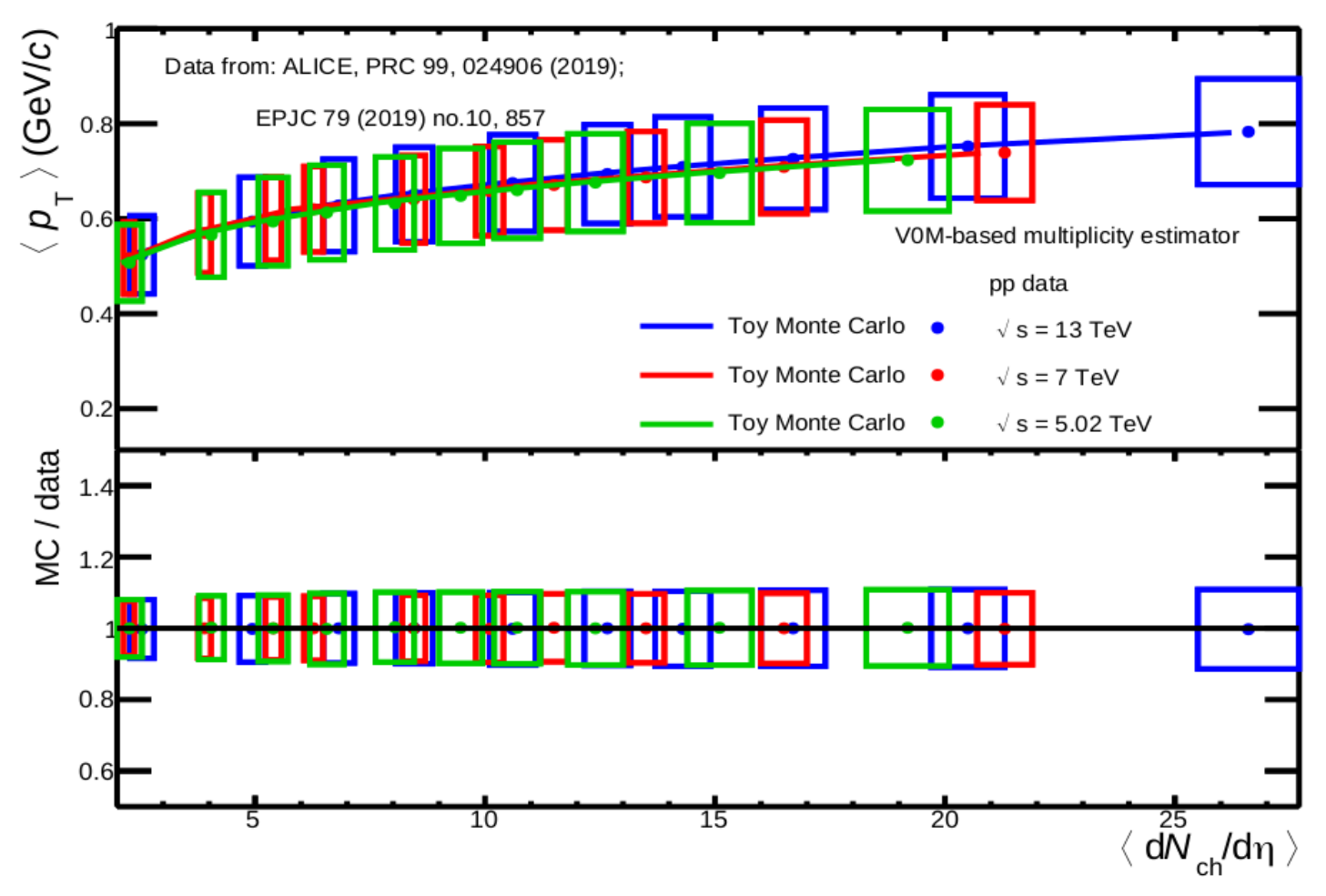}
\caption{Mean transverse momentum as a function of the average charged-particle multiplicity density in pp collisions at $\sqrt{s}=5.02$, 7 and 13\,TeV. In the top panel ALICE data~\cite{Acharya:2019mzb,Acharya:2018orn} (solid markers) are compared with results from a Toy Monte Carlo (solid lines). Bottom panel displays ratios between Toy Monte Carlo and the data. In all panels boxes around solid markers indicate the systematic uncertainties.}
\label{meanptvsN}  
\end{figure}

For simplicity, each event class was simulated assuming that its multiplicity spectrum follows a Poisson distribution~\cite{Golokhvastov:1994va}. Their corresponding average multiplicity values as well as their contribution to the inelastic cross section were taken from~\cite{Acharya:2018orn,Acharya:2019mzb}. With this information, \nch pseudo-particles were generated in each event, where each psuedo-particle had a transverse momentum which obeyed the \pt spectra reported by ALICE~\cite{Acharya:2018orn,Acharya:2019mzb}. The information of all events generated with the toy MC  was stored as a columnar dataset (TTree) using ROOT~\cite{Antcheva:2009zz}.  Figure~\ref{meanptvsN} displays the mean transverse momentum as a function of the average charged-particle multiplicity density in pp collisions at $\sqrt{s}=5.02$, 7 and 13\,TeV. The comparison between the toy MC and the data is displayed. Within uncertainties, the toy MC reproduces the correlation between the $\langle p_{\rm T} \rangle$ and $\langle d \nch/d\eta \rangle$. In our approach, the event-by-event information produced by the toy MC was processed with the trained BDT in order to extract the MPI activity associated with the data.

The toy MC approach was validated using  Pythia~8. A toy MC was built using the \pt spectra as a function of \nch obtained from Pythia~8. The MC non-closure (\nmpi from regression compared to the true \nmpi) was found to be significantly smaller than the systematic uncertainty due to model dependence. In addition, different conditions were varied to estimate a systematic uncertainty on the target variable. Fixing the spectral shape of the transverse momentum distribution, we vary the average charged-particle multiplicity density at their minimum and maximum values given by their corresponding uncertainties. On the other hand, we vary the average transverse momentum in the same way as multiplicity, but fixing the average charged-particle multiplicity density at their mean value. These variations provide an additional source of systematic uncertainty in our target variable, however, their contributions are also negligible with respect to the one due to the model dependence discussed before.

\section{Results}

\begin{figure}[t]
\includegraphics[width=0.45\textwidth]{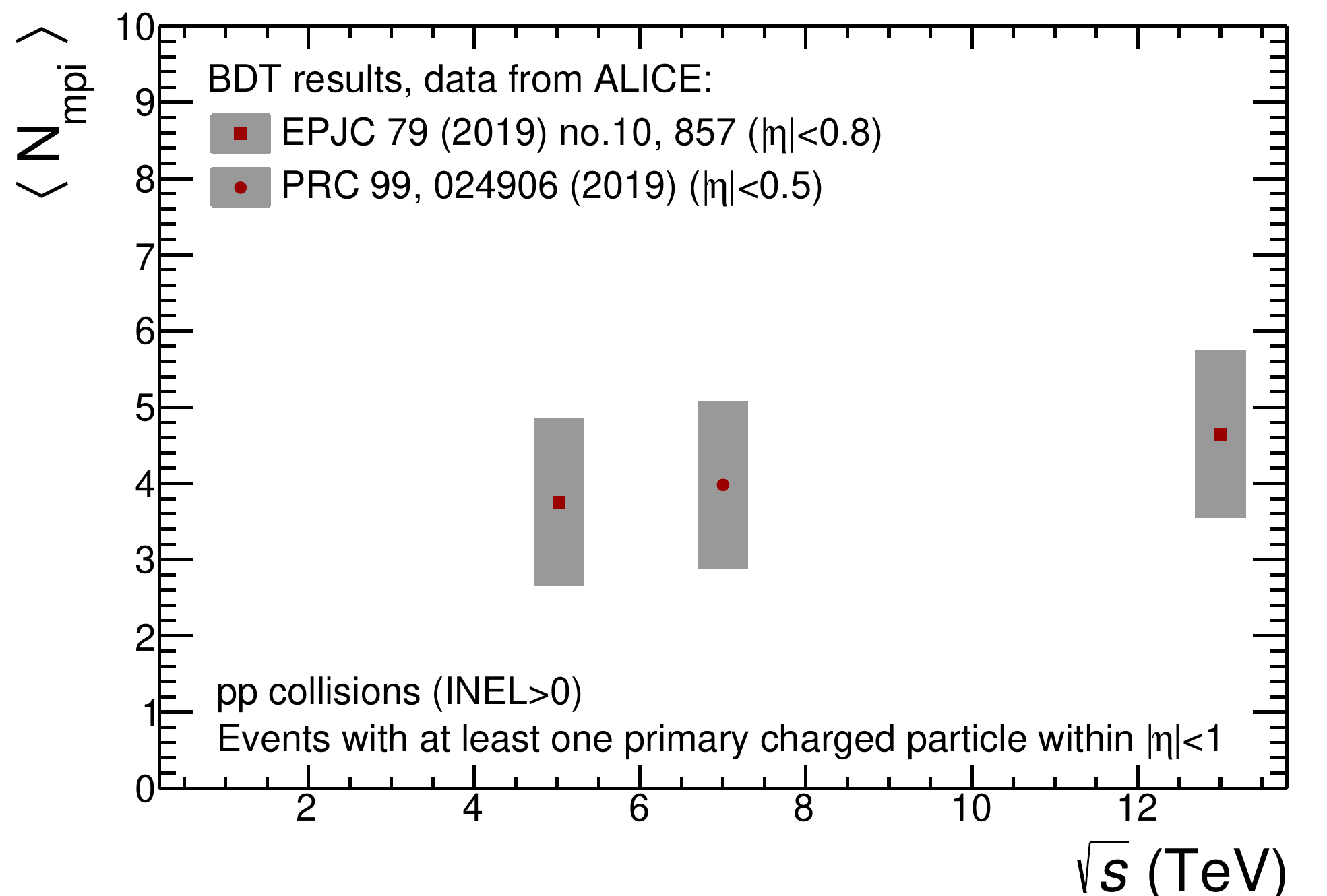}
\caption{Average number of MPI as a function of the center-of-
mass energy. The trained BDT were applied to ALICE data~\cite{Acharya:2018orn,Acharya:2019mzb}. Results for pp collisions at $\sqrt{s}=7$\,TeV, are compared to those for pp collisions at $\sqrt{s}=5.02$ and 13\,TeV reported in~\cite{Ortiz:2020rwg}.}
\label{fig:2}
\end{figure}

\begin{figure}[t]
\includegraphics[width=0.45\textwidth]{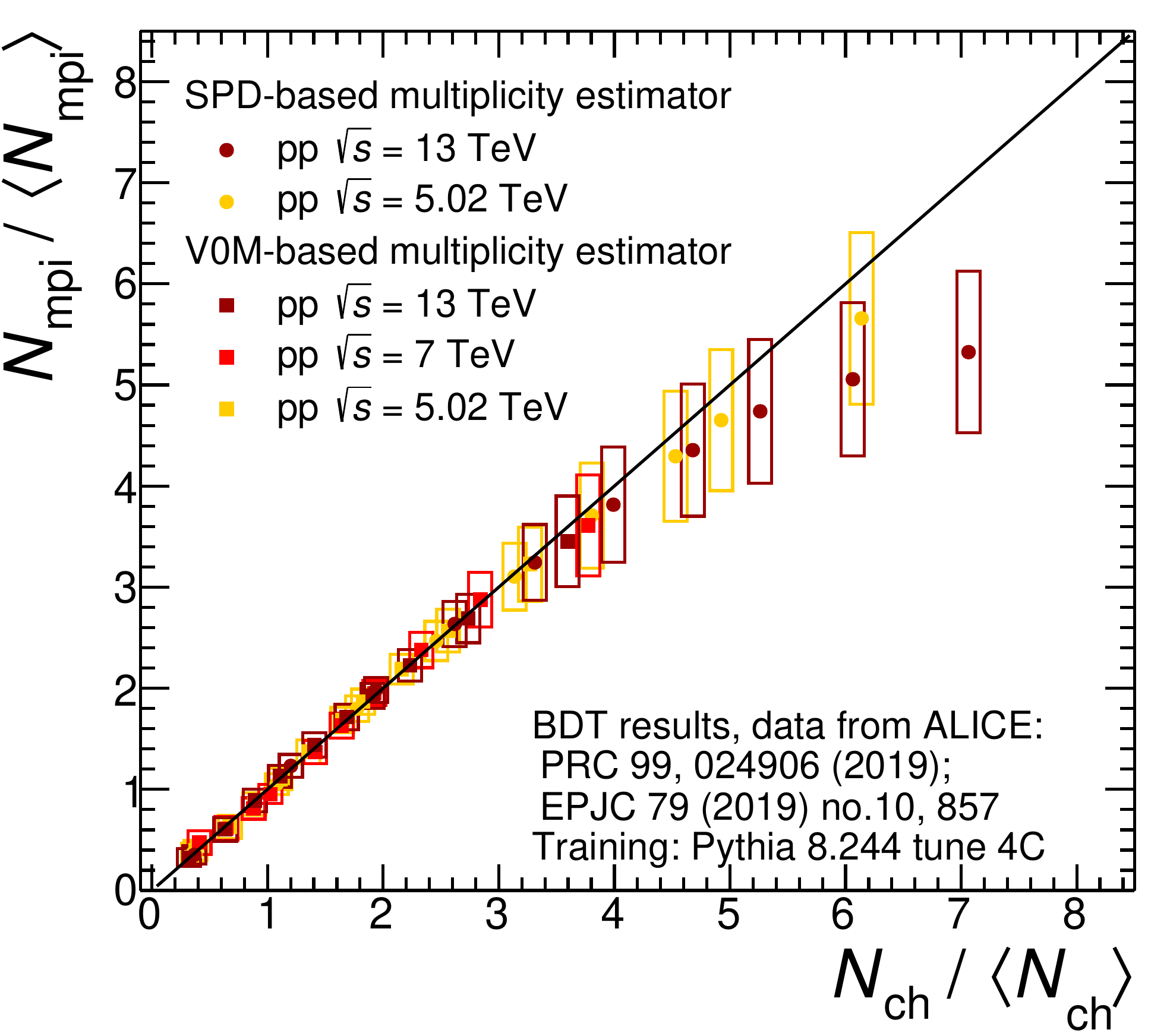}
\caption{The self normalized average number of Multiparton Interactions as a function of the self normalized mid-pseudorapidity charged particle multiplicity is shown for pp collisions at $\sqrt{s}=5.02$, 7 and 13\,TeV. The color boxes around the MC prediction indicate the systematic uncertainties (see the text for more details). Details about the V0M- and SPD-based estimators can be found in the text.}
\label{fig:3}  
\end{figure}

Firstly, we report a result which complements those reported in Ref.~\cite{Ortiz:2020rwg}. Following the same strategy discussed in Ref.~\cite{Ortiz:2020rwg}, we use the ALICE data from pp collisions at $\sqrt{s}=7$\,TeV~\cite{Acharya:2018orn} to get the average MPI activity. The average number of Multiparton Interactions is found to be $\langle N_{\rm mpi} \rangle = 3.98 \pm 1.01$, which is slightly smaller (higher) than that for pp collisions at $\sqrt{s}=13$\,TeV ($\sqrt{s}=5.02$\,TeV). It is worth mentioning that given the analysis strategy, the systematic uncertainties discussed here are expected to be correlated between \nch bins and energies.  Figure~\ref{fig:2} displays the average number of MPI as a function of the center-of-mass energy, for pp collision at $\sqrt{s}$= 5.02, $\sqrt{s}$= 7 TeV \cite{Acharya:2018orn}, and 13 TeV \cite{Acharya:2019mzb}. Within $3\sigma$, we obtain a regression value which is above unity, therefore, our results support the presence of MPI in pp collisions. We also observe a modest energy dependence, which is similar to that predicted by Pythia~8~\cite{Ortiz:2020rwg}. Secondly, figure~\ref{fig:3} displays the self normalized number of MPI  ($\nmpi / \langle \nmpi \rangle$) as a function of the self-normalized mid-pseudorapidity charged-particle multiplicity ($\nch / \langle \nch \rangle$) in pp collisions at $\sqrt{s}=5.02$, 7 and 13\,TeV from ALICE data. We observe that $\nmpi / \langle \nmpi \rangle$ vs. $\nch / \langle \nch \rangle$ does not show a significant center-of-mass energy dependence. Moreover, for $\nch  <3 \langle \nch \rangle $  the self normalized \nmpi increases linearly with the event multiplicity. While, for higher multiplicities, we observe a deviation of the self normalized \nmpi with respect to the linear trend. This result qualitatively agrees with Pythia~8 (see figure~\ref{fig:1}).

Last but not least, it is worth mentioning how our results compare with existing ALICE measurements sensitive to MPI~\cite{Abelev:2013sqa}. That analysis consists on the measurement of pair-yields per trigger in two-particle azimuthal correlations between charged trigger and associated particles in pp collisions at $\sqrt{s}=0.9$, 2.76 and 7\,TeV. The analysis was performed at mid-pseudorapidity ($|\eta|<0.9$) for the transverse momentum thresholds for trigger particles of $\pt^{\rm trigg.} > 0.7$\,GeV/$c$ and for associated particles of $\pt^{\rm assoc.} > 0.4$ and $0.7$\,GeV/$c$. Based on Pythia simulations, the so-called number of uncorrelated seeds is defined, and the results from data are discussed in the context of the semi-hard parton–parton interactions. The data indicate that the charged particle multiplicity increases approximately linearly with the number of uncorrelated seeds. However, it deviates from the linear dependence at large charged particle multiplicities. In addition, the data exhibit a weak center-of-mass energy dependence. These observations are fully consistent with our results which use Machine Learning. And they suggest that at highest multiplicities (at mid-pseudorapidity) a further increase of the number of Multiparton Interactions becomes very improbable, instead high multiplicities can only be reached by selecting events with many high-multiplicity jets~\cite{Abelev:2013sqa}. A similar conclusion is obtained from a study of the jet production as a function of event multiplicity in pp collisions~\cite{Chatrchyan:2013ala,Ortiz:2016kpz}.

\section{Conclusions}

In this work, we report the extraction of the average number of Multiparton Interactions from pp data at the LHC energies. Using the existing data on \pt spectra as a function of event multiplicity in pp collisions at $\sqrt{s}=7$\,TeV, we have found $\langle N_{\rm mpi} \rangle = 3.98 \pm 1.01$ for minimum-bias pp collisions. The comparisons with our previous results for pp collisions at $\sqrt{s}=5.02$ and 13\,TeV indicate a modest energy dependence of \nmpi. This observation is consistent with predictions by Pythia~8.244. Implicitly, our results also provide experimental evidence of the presence of MPI in hadronic interactions.  In addition, we also report the multiplicity dependence of \nmpi for the three center-of-mass energies. We have found that for $\nch  <3 \langle \nch \rangle $  the event multiplicity increases linearly with the self normalized \nmpi. While, for $\nch > 3 \langle \nch \rangle $, a deviation with respect to the linear trend of the self normalized \nmpi as a function of event multiplicity is observed. This suggests that these collisions can only be reached by
selecting events with high multiplicity
jets. All the results reported in this paper, are fully consistent with existing ALICE measurements, where a quantity sensitive to MPI was measured as a function of multiplicity and the center-of-mass energy. Based on all the crosschecks which were performed using MC, and the agreement with an independent measurement of ALICE at lower center-of-mass energies, the present results confirm that our approach is robust. Therefore, it can be used by experiments in order to study the particle production as a function of MPI. This will help to rule out models, and would contribute to the understanding of the heavy-ion-like features observed in pp data.

\begin{acknowledgments}
Authors acknowledge Antonio Paz for providing the simulations with Herwig~7.2. Support for this work has been received from CONACyT under the Grant No. A1-S-22917. E. Z. acknowledges the fellowship of CONACyT.
\end{acknowledgments}


\bibliography{biblio}

\begin{thebibliography}{35}%
\makeatletter
\providecommand \@ifxundefined [1]{%
 \@ifx{#1\undefined}
}%
\providecommand \@ifnum [1]{%
 \ifnum #1\expandafter \@firstoftwo
 \else \expandafter \@secondoftwo
 \fi
}%
\providecommand \@ifx [1]{%
 \ifx #1\expandafter \@firstoftwo
 \else \expandafter \@secondoftwo
 \fi
}%
\providecommand \natexlab [1]{#1}%
\providecommand \enquote  [1]{``#1''}%
\providecommand \bibnamefont  [1]{#1}%
\providecommand \bibfnamefont [1]{#1}%
\providecommand \citenamefont [1]{#1}%
\providecommand \href@noop [0]{\@secondoftwo}%
\providecommand \href [0]{\begingroup \@sanitize@url \@href}%
\providecommand \@href[1]{\@@startlink{#1}\@@href}%
\providecommand \@@href[1]{\endgroup#1\@@endlink}%
\providecommand \@sanitize@url [0]{\catcode `\\12\catcode `\$12\catcode
  `\&12\catcode `\#12\catcode `\^12\catcode `\_12\catcode `\%12\relax}%
\providecommand \@@startlink[1]{}%
\providecommand \@@endlink[0]{}%
\providecommand \url  [0]{\begingroup\@sanitize@url \@url }%
\providecommand \@url [1]{\endgroup\@href {#1}{\urlprefix }}%
\providecommand \urlprefix  [0]{URL }%
\providecommand \Eprint [0]{\href }%
\providecommand \doibase [0]{http://dx.doi.org/}%
\providecommand \selectlanguage [0]{\@gobble}%
\providecommand \bibinfo  [0]{\@secondoftwo}%
\providecommand \bibfield  [0]{\@secondoftwo}%
\providecommand \translation [1]{[#1]}%
\providecommand \BibitemOpen [0]{}%
\providecommand \bibitemStop [0]{}%
\providecommand \bibitemNoStop [0]{.\EOS\space}%
\providecommand \EOS [0]{\spacefactor3000\relax}%
\providecommand \BibitemShut  [1]{\csname bibitem#1\endcsname}%
\let\auto@bib@innerbib\@empty
\bibitem [{\citenamefont {Bahr}\ \emph {et~al.}(2009)\citenamefont {Bahr},
  \citenamefont {Butterworth},\ and\ \citenamefont {Seymour}}]{Bahr:2008wk}%
  \BibitemOpen
  \bibfield  {author} {\bibinfo {author} {\bibfnamefont {M.}~\bibnamefont
  {Bahr}}, \bibinfo {author} {\bibfnamefont {J.~M.}\ \bibnamefont
  {Butterworth}}, \ and\ \bibinfo {author} {\bibfnamefont {M.~H.}\ \bibnamefont
  {Seymour}},\ }\href {\doibase 10.1088/1126-6708/2009/01/065} {\bibfield
  {journal} {\bibinfo  {journal} {JHEP}\ }\textbf {\bibinfo {volume} {01}},\
  \bibinfo {pages} {065} (\bibinfo {year} {2009})},\ \Eprint
  {http://arxiv.org/abs/0806.2949} {arXiv:0806.2949 [hep-ph]} \BibitemShut
  {NoStop}%
\bibitem [{\citenamefont {Sj\"ostrand}\ and\ \citenamefont {van
  Zijl}(1987)}]{Sjostrand:1986ep}%
  \BibitemOpen
  \bibfield  {author} {\bibinfo {author} {\bibfnamefont {T.}~\bibnamefont
  {Sj\"ostrand}}\ and\ \bibinfo {author} {\bibfnamefont {M.}~\bibnamefont {van
  Zijl}},\ }\href {\doibase 10.1016/0370-2693(87)90722-2} {\bibfield  {journal}
  {\bibinfo  {journal} {Phys. Lett. B}\ }\textbf {\bibinfo {volume} {188}},\
  \bibinfo {pages} {149} (\bibinfo {year} {1987})}\BibitemShut {NoStop}%
\bibitem [{\citenamefont {Abelev}\ \emph {et~al.}(2012)\citenamefont {Abelev}
  \emph {et~al.}}]{Abelev:2012sk}%
  \BibitemOpen
  \bibfield  {author} {\bibinfo {author} {\bibfnamefont {B.}~\bibnamefont
  {Abelev}} \emph {et~al.} (\bibinfo {collaboration} {ALICE}),\ }\href
  {\doibase 10.1140/epjc/s10052-012-2124-9} {\bibfield  {journal} {\bibinfo
  {journal} {Eur. Phys. J. C}\ }\textbf {\bibinfo {volume} {72}},\ \bibinfo
  {pages} {2124} (\bibinfo {year} {2012})},\ \Eprint
  {http://arxiv.org/abs/1205.3963} {arXiv:1205.3963 [hep-ex]} \BibitemShut
  {NoStop}%
\bibitem [{\citenamefont {Abelev}\ \emph {et~al.}(2013)\citenamefont {Abelev}
  \emph {et~al.}}]{Abelev:2013sqa}%
  \BibitemOpen
  \bibfield  {author} {\bibinfo {author} {\bibfnamefont {B.}~\bibnamefont
  {Abelev}} \emph {et~al.} (\bibinfo {collaboration} {ALICE}),\ }\href
  {\doibase 10.1007/JHEP09(2013)049} {\bibfield  {journal} {\bibinfo  {journal}
  {JHEP}\ }\textbf {\bibinfo {volume} {09}},\ \bibinfo {pages} {049} (\bibinfo
  {year} {2013})},\ \Eprint {http://arxiv.org/abs/1307.1249} {arXiv:1307.1249
  [nucl-ex]} \BibitemShut {NoStop}%
\bibitem [{\citenamefont {Ortiz}(2018)}]{Ortiz:2017jho}%
  \BibitemOpen
  \bibfield  {author} {\bibinfo {author} {\bibfnamefont {A.}~\bibnamefont
  {Ortiz}},\ }\href {\doibase 10.1142/9789813227767_0016} {\bibfield  {journal}
  {\bibinfo  {journal} {Adv. Ser. Direct. High Energy Phys.}\ }\textbf
  {\bibinfo {volume} {29}},\ \bibinfo {pages} {343} (\bibinfo {year} {2018})},\
  \Eprint {http://arxiv.org/abs/1705.02056} {arXiv:1705.02056 [hep-ex]}
  \BibitemShut {NoStop}%
\bibitem [{\citenamefont {Ortiz}\ \emph {et~al.}(2020)\citenamefont {Ortiz},
  \citenamefont {Paz}, \citenamefont {Romo}, \citenamefont {Tripathy},
  \citenamefont {Zepeda},\ and\ \citenamefont {Bautista}}]{Ortiz:2020rwg}%
  \BibitemOpen
  \bibfield  {author} {\bibinfo {author} {\bibfnamefont {A.}~\bibnamefont
  {Ortiz}}, \bibinfo {author} {\bibfnamefont {A.}~\bibnamefont {Paz}}, \bibinfo
  {author} {\bibfnamefont {J.~D.}\ \bibnamefont {Romo}}, \bibinfo {author}
  {\bibfnamefont {S.}~\bibnamefont {Tripathy}}, \bibinfo {author}
  {\bibfnamefont {E.~A.}\ \bibnamefont {Zepeda}}, \ and\ \bibinfo {author}
  {\bibfnamefont {I.}~\bibnamefont {Bautista}},\ }\href {\doibase
  10.1103/PhysRevD.102.076014} {\bibfield  {journal} {\bibinfo  {journal}
  {Phys. Rev. D}\ }\textbf {\bibinfo {volume} {102}},\ \bibinfo {pages}
  {076014} (\bibinfo {year} {2020})},\ \Eprint
  {http://arxiv.org/abs/2004.03800} {arXiv:2004.03800 [hep-ph]} \BibitemShut
  {NoStop}%
\bibitem [{\citenamefont {Koba}\ \emph {et~al.}(1972)\citenamefont {Koba},
  \citenamefont {Nielsen},\ and\ \citenamefont {Olesen}}]{Koba:1972ng}%
  \BibitemOpen
  \bibfield  {author} {\bibinfo {author} {\bibfnamefont {Z.}~\bibnamefont
  {Koba}}, \bibinfo {author} {\bibfnamefont {H.~B.}\ \bibnamefont {Nielsen}}, \
  and\ \bibinfo {author} {\bibfnamefont {P.}~\bibnamefont {Olesen}},\ }\href
  {\doibase 10.1016/0550-3213(72)90551-2} {\bibfield  {journal} {\bibinfo
  {journal} {Nucl. Phys. B}\ }\textbf {\bibinfo {volume} {40}},\ \bibinfo
  {pages} {317} (\bibinfo {year} {1972})}\BibitemShut {NoStop}%
\bibitem [{\citenamefont {Alner}\ \emph {et~al.}(1986)\citenamefont {Alner}
  \emph {et~al.}}]{Alner:1985wj}%
  \BibitemOpen
  \bibfield  {author} {\bibinfo {author} {\bibfnamefont {G.}~\bibnamefont
  {Alner}} \emph {et~al.} (\bibinfo {collaboration} {UA5}),\ }\href {\doibase
  10.1016/0370-2693(86)91304-3} {\bibfield  {journal} {\bibinfo  {journal}
  {Phys. Lett. B}\ }\textbf {\bibinfo {volume} {167}},\ \bibinfo {pages} {476}
  (\bibinfo {year} {1986})}\BibitemShut {NoStop}%
\bibitem [{\citenamefont {Dremin}\ and\ \citenamefont
  {Nechitailo}(2011)}]{Dremin:2011sa}%
  \BibitemOpen
  \bibfield  {author} {\bibinfo {author} {\bibfnamefont {I.}~\bibnamefont
  {Dremin}}\ and\ \bibinfo {author} {\bibfnamefont {V.}~\bibnamefont
  {Nechitailo}},\ }\href {\doibase 10.1103/PhysRevD.84.034026} {\bibfield
  {journal} {\bibinfo  {journal} {Phys. Rev. D}\ }\textbf {\bibinfo {volume}
  {84}},\ \bibinfo {pages} {034026} (\bibinfo {year} {2011})},\ \Eprint
  {http://arxiv.org/abs/1106.4959} {arXiv:1106.4959 [hep-ph]} \BibitemShut
  {NoStop}%
\bibitem [{\citenamefont {Blok}\ \emph {et~al.}(2017)\citenamefont {Blok},
  \citenamefont {J\"akel}, \citenamefont {Strikman},\ and\ \citenamefont
  {Wiedemann}}]{Blok:2017pui}%
  \BibitemOpen
  \bibfield  {author} {\bibinfo {author} {\bibfnamefont {B.}~\bibnamefont
  {Blok}}, \bibinfo {author} {\bibfnamefont {C.~D.}\ \bibnamefont {J\"akel}},
  \bibinfo {author} {\bibfnamefont {M.}~\bibnamefont {Strikman}}, \ and\
  \bibinfo {author} {\bibfnamefont {U.~A.}\ \bibnamefont {Wiedemann}},\ }\href
  {\doibase 10.1007/JHEP12(2017)074} {\bibfield  {journal} {\bibinfo  {journal}
  {JHEP}\ }\textbf {\bibinfo {volume} {12}},\ \bibinfo {pages} {074} (\bibinfo
  {year} {2017})},\ \Eprint {http://arxiv.org/abs/1708.08241} {arXiv:1708.08241
  [hep-ph]} \BibitemShut {NoStop}%
\bibitem [{\citenamefont {Blok}\ and\ \citenamefont
  {Wiedemann}(2019)}]{Blok:2018xes}%
  \BibitemOpen
  \bibfield  {author} {\bibinfo {author} {\bibfnamefont {B.}~\bibnamefont
  {Blok}}\ and\ \bibinfo {author} {\bibfnamefont {U.~A.}\ \bibnamefont
  {Wiedemann}},\ }\href {\doibase 10.1016/j.physletb.2019.05.038} {\bibfield
  {journal} {\bibinfo  {journal} {Phys. Lett. B}\ }\textbf {\bibinfo {volume}
  {795}},\ \bibinfo {pages} {259} (\bibinfo {year} {2019})},\ \Eprint
  {http://arxiv.org/abs/1812.04113} {arXiv:1812.04113 [hep-ph]} \BibitemShut
  {NoStop}%
\bibitem [{\citenamefont {Khachatryan}\ \emph {et~al.}(2010)\citenamefont
  {Khachatryan} \emph {et~al.}}]{Khachatryan:2010gv}%
  \BibitemOpen
  \bibfield  {author} {\bibinfo {author} {\bibfnamefont {V.}~\bibnamefont
  {Khachatryan}} \emph {et~al.} (\bibinfo {collaboration} {CMS}),\ }\href
  {\doibase 10.1007/JHEP09(2010)091} {\bibfield  {journal} {\bibinfo  {journal}
  {JHEP}\ }\textbf {\bibinfo {volume} {09}},\ \bibinfo {pages} {091} (\bibinfo
  {year} {2010})},\ \Eprint {http://arxiv.org/abs/1009.4122} {arXiv:1009.4122
  [hep-ex]} \BibitemShut {NoStop}%
\bibitem [{\citenamefont {Adam}\ \emph {et~al.}(2017)\citenamefont {Adam} \emph
  {et~al.}}]{ALICE:2017jyt}%
  \BibitemOpen
  \bibfield  {author} {\bibinfo {author} {\bibfnamefont {J.}~\bibnamefont
  {Adam}} \emph {et~al.} (\bibinfo {collaboration} {ALICE}),\ }\href {\doibase
  10.1038/nphys4111} {\bibfield  {journal} {\bibinfo  {journal} {Nature Phys.}\
  }\textbf {\bibinfo {volume} {13}},\ \bibinfo {pages} {535} (\bibinfo {year}
  {2017})},\ \Eprint {http://arxiv.org/abs/1606.07424} {arXiv:1606.07424
  [nucl-ex]} \BibitemShut {NoStop}%
\bibitem [{\citenamefont {Acharya}\ \emph
  {et~al.}(2019{\natexlab{a}})\citenamefont {Acharya} \emph
  {et~al.}}]{Acharya:2018orn}%
  \BibitemOpen
  \bibfield  {author} {\bibinfo {author} {\bibfnamefont {S.}~\bibnamefont
  {Acharya}} \emph {et~al.} (\bibinfo {collaboration} {ALICE}),\ }\href
  {\doibase 10.1103/PhysRevC.99.024906} {\bibfield  {journal} {\bibinfo
  {journal} {Phys. Rev. C}\ }\textbf {\bibinfo {volume} {99}},\ \bibinfo
  {pages} {024906} (\bibinfo {year} {2019}{\natexlab{a}})},\ \Eprint
  {http://arxiv.org/abs/1807.11321} {arXiv:1807.11321 [nucl-ex]} \BibitemShut
  {NoStop}%
\bibitem [{\citenamefont {Bozek}(2012)}]{Bozek:2011if}%
  \BibitemOpen
  \bibfield  {author} {\bibinfo {author} {\bibfnamefont {P.}~\bibnamefont
  {Bozek}},\ }\href {\doibase 10.1103/PhysRevC.85.014911} {\bibfield  {journal}
  {\bibinfo  {journal} {Phys. Rev.}\ }\textbf {\bibinfo {volume} {C85}},\
  \bibinfo {pages} {014911} (\bibinfo {year} {2012})},\ \Eprint
  {http://arxiv.org/abs/1112.0915} {arXiv:1112.0915 [hep-ph]} \BibitemShut
  {NoStop}%
\bibitem [{\citenamefont {Nagle}\ and\ \citenamefont
  {Zajc}(2018)}]{Nagle:2018nvi}%
  \BibitemOpen
  \bibfield  {author} {\bibinfo {author} {\bibfnamefont {J.~L.}\ \bibnamefont
  {Nagle}}\ and\ \bibinfo {author} {\bibfnamefont {W.~A.}\ \bibnamefont
  {Zajc}},\ }\href {\doibase 10.1146/annurev-nucl-101916-123209} {\bibfield
  {journal} {\bibinfo  {journal} {Ann. Rev. Nucl. Part. Sci.}\ }\textbf
  {\bibinfo {volume} {68}},\ \bibinfo {pages} {211} (\bibinfo {year} {2018})},\
  \Eprint {http://arxiv.org/abs/1801.03477} {arXiv:1801.03477 [nucl-ex]}
  \BibitemShut {NoStop}%
\bibitem [{\citenamefont {Sj\"ostrand}\ \emph {et~al.}(2015)\citenamefont
  {Sj\"ostrand}, \citenamefont {Ask}, \citenamefont {Christiansen},
  \citenamefont {Corke}, \citenamefont {Desai}, \citenamefont {Ilten},
  \citenamefont {Mrenna}, \citenamefont {Prestel}, \citenamefont {Rasmussen},\
  and\ \citenamefont {Skands}}]{Sjostrand:2014zea}%
  \BibitemOpen
  \bibfield  {author} {\bibinfo {author} {\bibfnamefont {T.}~\bibnamefont
  {Sj\"ostrand}}, \bibinfo {author} {\bibfnamefont {S.}~\bibnamefont {Ask}},
  \bibinfo {author} {\bibfnamefont {J.~R.}\ \bibnamefont {Christiansen}},
  \bibinfo {author} {\bibfnamefont {R.}~\bibnamefont {Corke}}, \bibinfo
  {author} {\bibfnamefont {N.}~\bibnamefont {Desai}}, \bibinfo {author}
  {\bibfnamefont {P.}~\bibnamefont {Ilten}}, \bibinfo {author} {\bibfnamefont
  {S.}~\bibnamefont {Mrenna}}, \bibinfo {author} {\bibfnamefont
  {S.}~\bibnamefont {Prestel}}, \bibinfo {author} {\bibfnamefont {C.~O.}\
  \bibnamefont {Rasmussen}}, \ and\ \bibinfo {author} {\bibfnamefont {P.~Z.}\
  \bibnamefont {Skands}},\ }\href {\doibase 10.1016/j.cpc.2015.01.024}
  {\bibfield  {journal} {\bibinfo  {journal} {Comput. Phys. Commun.}\ }\textbf
  {\bibinfo {volume} {191}},\ \bibinfo {pages} {159} (\bibinfo {year}
  {2015})},\ \Eprint {http://arxiv.org/abs/1410.3012} {arXiv:1410.3012
  [hep-ph]} \BibitemShut {NoStop}%
\bibitem [{\citenamefont {Gieseke}\ \emph {et~al.}(2012)\citenamefont
  {Gieseke}, \citenamefont {Rohr},\ and\ \citenamefont
  {Siodmok}}]{Gieseke:2012ft}%
  \BibitemOpen
  \bibfield  {author} {\bibinfo {author} {\bibfnamefont {S.}~\bibnamefont
  {Gieseke}}, \bibinfo {author} {\bibfnamefont {C.}~\bibnamefont {Rohr}}, \
  and\ \bibinfo {author} {\bibfnamefont {A.}~\bibnamefont {Siodmok}},\ }\href
  {\doibase 10.1140/epjc/s10052-012-2225-5} {\bibfield  {journal} {\bibinfo
  {journal} {Eur. Phys. J. C}\ }\textbf {\bibinfo {volume} {72}},\ \bibinfo
  {pages} {2225} (\bibinfo {year} {2012})},\ \Eprint
  {http://arxiv.org/abs/1206.0041} {arXiv:1206.0041 [hep-ph]} \BibitemShut
  {NoStop}%
\bibitem [{\citenamefont {Ortiz}\ \emph {et~al.}(2013)\citenamefont {Ortiz},
  \citenamefont {Christiansen}, \citenamefont {Cuautle~Flores}, \citenamefont
  {Maldonado~Cervantes},\ and\ \citenamefont {Pai\'c}}]{Ortiz:2013yxa}%
  \BibitemOpen
  \bibfield  {author} {\bibinfo {author} {\bibfnamefont {A.}~\bibnamefont
  {Ortiz}}, \bibinfo {author} {\bibfnamefont {P.}~\bibnamefont {Christiansen}},
  \bibinfo {author} {\bibfnamefont {E.}~\bibnamefont {Cuautle~Flores}},
  \bibinfo {author} {\bibfnamefont {I.}~\bibnamefont {Maldonado~Cervantes}}, \
  and\ \bibinfo {author} {\bibfnamefont {G.}~\bibnamefont {Pai\'c}},\ }\href
  {\doibase 10.1103/PhysRevLett.111.042001} {\bibfield  {journal} {\bibinfo
  {journal} {Phys. Rev. Lett.}\ }\textbf {\bibinfo {volume} {111}},\ \bibinfo
  {pages} {042001} (\bibinfo {year} {2013})},\ \Eprint
  {http://arxiv.org/abs/1303.6326} {arXiv:1303.6326 [hep-ph]} \BibitemShut
  {NoStop}%
\bibitem [{\citenamefont {Bierlich}\ and\ \citenamefont
  {Christiansen}(2015)}]{Bierlich:2015rha}%
  \BibitemOpen
  \bibfield  {author} {\bibinfo {author} {\bibfnamefont {C.}~\bibnamefont
  {Bierlich}}\ and\ \bibinfo {author} {\bibfnamefont {J.~R.}\ \bibnamefont
  {Christiansen}},\ }\href {\doibase 10.1103/PhysRevD.92.094010} {\bibfield
  {journal} {\bibinfo  {journal} {Phys. Rev. D}\ }\textbf {\bibinfo {volume}
  {92}},\ \bibinfo {pages} {094010} (\bibinfo {year} {2015})},\ \Eprint
  {http://arxiv.org/abs/1507.02091} {arXiv:1507.02091 [hep-ph]} \BibitemShut
  {NoStop}%
\bibitem [{\citenamefont {Nayak}\ \emph {et~al.}(2019)\citenamefont {Nayak},
  \citenamefont {Pal},\ and\ \citenamefont {Dash}}]{Nayak:2018xip}%
  \BibitemOpen
  \bibfield  {author} {\bibinfo {author} {\bibfnamefont {R.}~\bibnamefont
  {Nayak}}, \bibinfo {author} {\bibfnamefont {S.}~\bibnamefont {Pal}}, \ and\
  \bibinfo {author} {\bibfnamefont {S.}~\bibnamefont {Dash}},\ }\href {\doibase
  10.1103/PhysRevD.100.074023} {\bibfield  {journal} {\bibinfo  {journal}
  {Phys. Rev. D}\ }\textbf {\bibinfo {volume} {100}},\ \bibinfo {pages}
  {074023} (\bibinfo {year} {2019})},\ \Eprint
  {http://arxiv.org/abs/1812.07718} {arXiv:1812.07718 [hep-ph]} \BibitemShut
  {NoStop}%
\bibitem [{\citenamefont {Mishra}\ \emph {et~al.}(2019)\citenamefont {Mishra},
  \citenamefont {Ortiz},\ and\ \citenamefont {Paic}}]{Mishra:2018pio}%
  \BibitemOpen
  \bibfield  {author} {\bibinfo {author} {\bibfnamefont {A.~N.}\ \bibnamefont
  {Mishra}}, \bibinfo {author} {\bibfnamefont {A.}~\bibnamefont {Ortiz}}, \
  and\ \bibinfo {author} {\bibfnamefont {G.}~\bibnamefont {Paic}},\ }\href
  {\doibase 10.1103/PhysRevC.99.034911} {\bibfield  {journal} {\bibinfo
  {journal} {Phys. Rev. C}\ }\textbf {\bibinfo {volume} {99}},\ \bibinfo
  {pages} {034911} (\bibinfo {year} {2019})},\ \Eprint
  {http://arxiv.org/abs/1805.04572} {arXiv:1805.04572 [hep-ph]} \BibitemShut
  {NoStop}%
\bibitem [{\citenamefont {Jacobs}(2021)}]{Jacobs:2020ptj}%
  \BibitemOpen
  \bibfield  {author} {\bibinfo {author} {\bibfnamefont {P.}~\bibnamefont
  {Jacobs}} (\bibinfo {collaboration} {ALICE}),\ }\href {\doibase
  10.1016/j.nuclphysa.2020.121924} {\bibfield  {journal} {\bibinfo  {journal}
  {Nucl. Phys. A}\ }\textbf {\bibinfo {volume} {1005}},\ \bibinfo {pages}
  {121924} (\bibinfo {year} {2021})},\ \Eprint
  {http://arxiv.org/abs/2001.09517} {arXiv:2001.09517 [nucl-ex]} \BibitemShut
  {NoStop}%
\bibitem [{\citenamefont {Benc\'edi}\ \emph {et~al.}(2020)\citenamefont
  {Benc\'edi}, \citenamefont {Ortiz},\ and\ \citenamefont
  {Tripathy}}]{Ortiz:2020dph}%
  \BibitemOpen
  \bibfield  {author} {\bibinfo {author} {\bibfnamefont {G.}~\bibnamefont
  {Benc\'edi}}, \bibinfo {author} {\bibfnamefont {A.}~\bibnamefont {Ortiz}}, \
  and\ \bibinfo {author} {\bibfnamefont {S.}~\bibnamefont {Tripathy}},\ }\href
  {\doibase 10.1088/1361-6471/abc5fb} {\bibfield  {journal} {\bibinfo
  {journal} {J. Phys. G}\ }\textbf {\bibinfo {volume} {48}},\ \bibinfo {pages}
  {015007} (\bibinfo {year} {2020})},\ \Eprint
  {http://arxiv.org/abs/2007.03857} {arXiv:2007.03857 [hep-ph]} \BibitemShut
  {NoStop}%
\bibitem [{\citenamefont {Acharya}\ \emph
  {et~al.}(2019{\natexlab{b}})\citenamefont {Acharya} \emph
  {et~al.}}]{Acharya:2019mzb}%
  \BibitemOpen
  \bibfield  {author} {\bibinfo {author} {\bibfnamefont {S.}~\bibnamefont
  {Acharya}} \emph {et~al.} (\bibinfo {collaboration} {ALICE}),\ }\href
  {\doibase 10.1140/epjc/s10052-019-7350-y} {\bibfield  {journal} {\bibinfo
  {journal} {Eur. Phys. J. C}\ }\textbf {\bibinfo {volume} {79}},\ \bibinfo
  {pages} {857} (\bibinfo {year} {2019}{\natexlab{b}})},\ \Eprint
  {http://arxiv.org/abs/1905.07208} {arXiv:1905.07208 [nucl-ex]} \BibitemShut
  {NoStop}%
\bibitem [{\citenamefont {Voss}\ \emph {et~al.}(2007)\citenamefont {Voss},
  \citenamefont {Hocker}, \citenamefont {Stelzer},\ and\ \citenamefont
  {Tegenfeldt}}]{Voss:2007jxm}%
  \BibitemOpen
  \bibfield  {author} {\bibinfo {author} {\bibfnamefont {H.}~\bibnamefont
  {Voss}}, \bibinfo {author} {\bibfnamefont {A.}~\bibnamefont {Hocker}},
  \bibinfo {author} {\bibfnamefont {J.}~\bibnamefont {Stelzer}}, \ and\
  \bibinfo {author} {\bibfnamefont {F.}~\bibnamefont {Tegenfeldt}},\ }\href
  {\doibase 10.22323/1.050.0040} {\bibfield  {journal} {\bibinfo  {journal}
  {PoS}\ }\textbf {\bibinfo {volume} {ACAT}},\ \bibinfo {pages} {040} (\bibinfo
  {year} {2007})}\BibitemShut {NoStop}%
\bibitem [{\citenamefont {Corke}\ and\ \citenamefont
  {Sjostrand}(2011)}]{Corke:2010yf}%
  \BibitemOpen
  \bibfield  {author} {\bibinfo {author} {\bibfnamefont {R.}~\bibnamefont
  {Corke}}\ and\ \bibinfo {author} {\bibfnamefont {T.}~\bibnamefont
  {Sjostrand}},\ }\href {\doibase 10.1007/JHEP03(2011)032} {\bibfield
  {journal} {\bibinfo  {journal} {JHEP}\ }\textbf {\bibinfo {volume} {03}},\
  \bibinfo {pages} {032} (\bibinfo {year} {2011})},\ \Eprint
  {http://arxiv.org/abs/1011.1759} {arXiv:1011.1759 [hep-ph]} \BibitemShut
  {NoStop}%
\bibitem [{\citenamefont {Cuautle}\ \emph {et~al.}(2016)\citenamefont
  {Cuautle}, \citenamefont {Ortiz},\ and\ \citenamefont
  {Paic}}]{Cuautle:2015fbx}%
  \BibitemOpen
  \bibfield  {author} {\bibinfo {author} {\bibfnamefont {E.}~\bibnamefont
  {Cuautle}}, \bibinfo {author} {\bibfnamefont {A.}~\bibnamefont {Ortiz}}, \
  and\ \bibinfo {author} {\bibfnamefont {G.}~\bibnamefont {Paic}},\ }\href
  {\doibase 10.1016/j.nuclphysa.2016.02.031} {\bibfield  {journal} {\bibinfo
  {journal} {Nucl. Phys. A}\ }\textbf {\bibinfo {volume} {956}},\ \bibinfo
  {pages} {749} (\bibinfo {year} {2016})},\ \Eprint
  {http://arxiv.org/abs/1512.09011} {arXiv:1512.09011 [hep-ph]} \BibitemShut
  {NoStop}%
\bibitem [{\citenamefont {Skands}\ \emph {et~al.}(2014)\citenamefont {Skands},
  \citenamefont {Carrazza},\ and\ \citenamefont {Rojo}}]{Skands:2014pea}%
  \BibitemOpen
  \bibfield  {author} {\bibinfo {author} {\bibfnamefont {P.}~\bibnamefont
  {Skands}}, \bibinfo {author} {\bibfnamefont {S.}~\bibnamefont {Carrazza}}, \
  and\ \bibinfo {author} {\bibfnamefont {J.}~\bibnamefont {Rojo}},\ }\href
  {\doibase 10.1140/epjc/s10052-014-3024-y} {\bibfield  {journal} {\bibinfo
  {journal} {Eur. Phys. J.}\ }\textbf {\bibinfo {volume} {C74}},\ \bibinfo
  {pages} {3024} (\bibinfo {year} {2014})},\ \Eprint
  {http://arxiv.org/abs/1404.5630} {arXiv:1404.5630 [hep-ph]} \BibitemShut
  {NoStop}%
\bibitem [{\citenamefont {Bellm}\ \emph {et~al.}(2020)\citenamefont {Bellm}
  \emph {et~al.}}]{Bellm:2019zci}%
  \BibitemOpen
  \bibfield  {author} {\bibinfo {author} {\bibfnamefont {J.}~\bibnamefont
  {Bellm}} \emph {et~al.},\ }\href {\doibase 10.1140/epjc/s10052-020-8011-x}
  {\bibfield  {journal} {\bibinfo  {journal} {Eur. Phys. J. C}\ }\textbf
  {\bibinfo {volume} {80}},\ \bibinfo {pages} {452} (\bibinfo {year} {2020})},\
  \Eprint {http://arxiv.org/abs/1912.06509} {arXiv:1912.06509 [hep-ph]}
  \BibitemShut {NoStop}%
\bibitem [{\citenamefont {Ortiz}\ \emph {et~al.}(2017)\citenamefont {Ortiz},
  \citenamefont {Bencedi},\ and\ \citenamefont {Bello}}]{Ortiz:2016kpz}%
  \BibitemOpen
  \bibfield  {author} {\bibinfo {author} {\bibfnamefont {A.}~\bibnamefont
  {Ortiz}}, \bibinfo {author} {\bibfnamefont {G.}~\bibnamefont {Bencedi}}, \
  and\ \bibinfo {author} {\bibfnamefont {H.}~\bibnamefont {Bello}},\ }\href
  {\doibase 10.1088/1361-6471/aa6594} {\bibfield  {journal} {\bibinfo
  {journal} {J. Phys. G}\ }\textbf {\bibinfo {volume} {44}},\ \bibinfo {pages}
  {065001} (\bibinfo {year} {2017})},\ \Eprint
  {http://arxiv.org/abs/1608.04784} {arXiv:1608.04784 [hep-ph]} \BibitemShut
  {NoStop}%
\bibitem [{doi(2014)}]{doi:10.1142/S0217751X14300440}%
  \BibitemOpen
  \href {\doibase 10.1142/S0217751X14300440} {\bibfield  {journal} {\bibinfo
  {journal} {International Journal of Modern Physics A}\ }\textbf {\bibinfo
  {volume} {29}},\ \bibinfo {pages} {1430044} (\bibinfo {year}
  {2014})}\BibitemShut {NoStop}%
\bibitem [{\citenamefont {Golokhvastov}(1995)}]{Golokhvastov:1994va}%
  \BibitemOpen
  \bibfield  {author} {\bibinfo {author} {\bibfnamefont {A.}~\bibnamefont
  {Golokhvastov}},\ }\href@noop {} {\bibfield  {journal} {\bibinfo  {journal}
  {Phys. Atom. Nucl.}\ }\textbf {\bibinfo {volume} {58}},\ \bibinfo {pages}
  {1998} (\bibinfo {year} {1995})}\BibitemShut {NoStop}%
\bibitem [{\citenamefont {Antcheva}\ \emph {et~al.}(2009)\citenamefont
  {Antcheva} \emph {et~al.}}]{Antcheva:2009zz}%
  \BibitemOpen
  \bibfield  {author} {\bibinfo {author} {\bibfnamefont {I.}~\bibnamefont
  {Antcheva}} \emph {et~al.},\ }\href {\doibase 10.1016/j.cpc.2009.08.005}
  {\bibfield  {journal} {\bibinfo  {journal} {Comput. Phys. Commun.}\ }\textbf
  {\bibinfo {volume} {180}},\ \bibinfo {pages} {2499} (\bibinfo {year}
  {2009})},\ \Eprint {http://arxiv.org/abs/1508.07749} {arXiv:1508.07749
  [physics.data-an]} \BibitemShut {NoStop}%
\bibitem [{\citenamefont {Chatrchyan}\ \emph {et~al.}(2013)\citenamefont
  {Chatrchyan} \emph {et~al.}}]{Chatrchyan:2013ala}%
  \BibitemOpen
  \bibfield  {author} {\bibinfo {author} {\bibfnamefont {S.}~\bibnamefont
  {Chatrchyan}} \emph {et~al.} (\bibinfo {collaboration} {CMS}),\ }\href
  {\doibase 10.1140/epjc/s10052-013-2674-5} {\bibfield  {journal} {\bibinfo
  {journal} {Eur. Phys. J. C}\ }\textbf {\bibinfo {volume} {73}},\ \bibinfo
  {pages} {2674} (\bibinfo {year} {2013})},\ \Eprint
  {http://arxiv.org/abs/1310.4554} {arXiv:1310.4554 [hep-ex]} \BibitemShut
  {NoStop}%
\end{thebibliography}%

\end{document}